\documentclass[runningheads,a4paper]{llncs}

\usepackage{amssymb,amsmath}
\usepackage{graphicx}

\usepackage{url}

\DeclareMathOperator{\supp}{supp}
\DeclareMathOperator{\dom}{dom}

\DeclareMathOperator{\ar}{ar}

\newcommand{\q}{\ensuremath{\mathbb{Q}}}

\newcommand{\fPol}[1]{\ensuremath{\operatorname{fPol}(#1)}}

\DeclareMathOperator{\pol}{Pol}

\DeclareMathOperator{\CostF}{{\bf \Phi}}

\newcommand{\CSP}[1]{\ensuremath{\operatorname{CSP}(#1)}}
\newcommand{\cspo}{\textsc{CSP}}
\newcommand{\vcspo}{\textsc{VCSP}}
\newcommand{\vcsp}[1]{\ensuremath{\operatorname{VCSP}(#1)}}
\newcommand{\VCSP}[1]{\ensuremath{\operatorname{VCSP}(#1)}}

\long\def\ignore#1{}
\def\myps[#1]#2{\includegraphics[#1]{#2}}

\def\br(#1,#2){{\langle #1,#2 \rangle}}
\def\setZ[#1,#2]{{[ #1 .. #2 ]}}

\newcommand{\Qc}{\mbox{$\overline{\mathbb Q}$}}

\def\q={\quad=\quad}
\def\qq={\qquad=\qquad}

\def\calI{{\cal I}}
\def\calH{{\cal H}}

\def\calO{{\cal O}}

\def\psfile[#1]#2{}
\def\psfilehere[#1]#2{}

\def\assign(#1,#2){\langle#1,#2\rangle}
\def\edge(#1,#2){(#1,#2)}

\def\slack(#1){\texttt{slack}({#1})}
\def\barslack(#1){\overline{\texttt{slack}}({#1})}

\def\unitvec(#1){{{\bf u}_{#1}}}

\newcommand{\bGamma}{\mathbf{\Gamma}}
\newcommand{\bR}{\mathbf{R}}
\newcommand{\bI}{\mathbf{I}}

\newcommand{\bv}{\mathbf{v}}

\newcommand{\up}{{\bf Up}}

\makeatletter
\def\thm@space@setup{%
  \thm@preskip=2pt \thm@postskip=2pt
}
\makeatother

\newif\ifTR\TRtrue
\newif\ifNO\NOfalse

\begin{document}

\mainmatter  

\title{Hybrid VCSPs with crisp and conservative valued templates}

\titlerunning{Hybrid VCSPs}

%
%
\author{Rustem Takhanov}
\authorrunning{Hybrid VCSPs}

\institute{Nazarbayev University,\\
rustem.takhanov@nu.edu.kz \\
}

%
%

\toctitle{Lecture Notes in Computer Science}
\tocauthor{Authors' Instructions}
\maketitle

\sloppy
\begin{abstract}
A constraint satisfaction problem (CSP) is a problem of computing a homomorphism
$\mbox{$\bR \rightarrow \bGamma$}$ between two relational structures, e.g.\ between two directed graphs.
Analyzing its complexity has been a very fruitful research direction, especially for {\em fixed template CSPs} (or, {\em non-uniform CSPs}), denoted $\textsc{CSP}(\bGamma)$,
in which the right side structure $\bGamma$ is fixed and the left side structure $\bR$ is unconstrained. 

Recently, the {\em hybrid} setting, written $\textsc{CSP}_{\mathcal{H}}(\bGamma)$, where both sides are restricted simultaneously, attracted some attention.
It assumes that $\bR$ is taken from a class of relational structures $\mathcal{H}$ (called the {\em structural restriction}) that additionally is {\em closed under inverse homomorphisms}. The last property allows to exploit an algebraic machinery that has been developed for fixed template CSPs. The key concept that connects hybrid CSPs with fixed-template CSPs is the so called ``lifted language''. Namely, this is a constraint language $\bGamma_{\bR}$ that can be constructed from an input $\bR$. The tractability of the language $\bGamma_{\bR}$ for any input $\bR\in\mathcal{H}$ is a necessary condition for the tractability of the hybrid problem.

In the first part we investigate templates $\bGamma$ for which the latter condition is not only necessary, but also is sufficient. We call such templates $\bGamma$ {\em widely tractable}. For this purpose, we construct from $\bGamma$ a new finite relational structure $\bGamma'$ and define a ``maximal'' structural restriction $\mathcal{H}_0$ as a class of structures homomorphic to $\bGamma'$.
For the so called strongly BJK templates that probably captures all templates, we prove that wide tractability is equivalent to the tractability of $\textsc{CSP}_{\mathcal{H}_0}(\bGamma)$.
Our proof is based on the key observation that $\bR$ is homomorphic to $\bGamma'$ if and only if the core of $\bGamma_{\bR}$ is preserved by a Siggers polymorphism.
Analogous result is shown for conservative valued CSPs.
 \end{abstract}

\section{Introduction}\label{sec:intro}
The {\it constraint satisfaction problems (CSPs)} and the {\it valued constraint satisfaction problems (VCSPs)} provide a powerful framework for the analysis of a large set of computational problems arising in propositional logic, combinatorial optimization, graph theory, artificial intelligence, scheduling, biology (protein folding), computer vision etc. CSP can be formalized either as a problem of (a) finding an assignment of values to a given set of variables, subject to constraints on the values that can be assigned simultaneously to specified subsets of variables, or as a problem of (b) finding a homomorphism between two finite relational structures $A$ and $B$ (e.g., two oriented graphs). 
These two formulations are polynomially equivalent under the condition that input constraints in the first case or input relations in the second case are given by lists of their elements. A soft version of CSP, the Valued CSP, generalizes the CSP by changing crisp constraints to cost functions applied to tuples of variables. In the VCSP we are asked to find a minimum (or maximum) of a sum of cost functions applied to corresponding variables. 

The CSPs have been a very active research field since 70s. One of the topics that revealed the 
rich logical and algebraic structure of the CSPs was the problem's computational complexity when constraint relations are restricted to a given set of relations or, alternatively, when the second relational structure is some fixed $\bGamma$. Thus, this problem is parameterized by $\bGamma$, denoted as $\CSP\bGamma$ and called a fixed template CSP with a template $\bGamma$ (another name is a non-uniform CSP).
E.g., if the domain set is boolean and $\bGamma$ is a structure with four ternary relations $x\vee y\vee z$,
$\overline{x}\vee y\vee z$, $\overline{x}\vee \overline{y}\vee z$,
$\overline{x}\vee \overline{y}\vee \overline{z}$, $\CSP\bGamma$ models 3-SAT which is historically one of the first NP-complete problems~\cite{Cook:1971}. At the same time, if we restrict $\bGamma$ to binary relations, then we obtain tractable 2-SAT. Schaeffer proved~\cite{schaefer78:complexity} that for any template $\bGamma$ over the boolean set,  $\CSP\bGamma$ is either in P or NP-complete. For the case when $\bGamma$ is a graph (without loops) Hell and Ne\v{s}et\v{r}il~\cite{Hell} proved an analogous statement, by showing that only for bipartite graphs the problem is tractable. Feder and Vardi~\cite{feder98:monotone} found that all fixed template CSPs can be expressed as problems in a fragment of SNP, called the Monotone Monadic SNP (MM SNP), and showed that for any problem in MM SNP there is a polynomial-time Turing reduction to a fixed template CSP. Thus, non-uniform CSPs' complexity classification would yield a classification for MM SNP problems. This result placed fixed-template CSPs into a broad logical context which naturally lead to a conjecture that such CSPs are either tractable or NP-hard, the so called dichotomy conjecture.

In \cite{Jeavons:1998} Jeavons showed that the complexity of $\CSP\bGamma$ is determined by the polymorphisms of $\bGamma$. Research in this direction lead to a conjectured description of tractable templates through properties of their polymorphisms. The key formulation was given by Bulatov, Jeavons, and Krokhin \cite{bulatov05:classifying}, with subsequent reformulations of this conjecture by Maroti and McKenzie \cite{MarotiMcKenzie}. Later, it was shown by Siggers~\cite{Siggers} that if the Bulatov-Jeavons-Krokhin formulation is true, then for a relational structure to be tractable it is necessary and sufficient that its core is preserved by a single 6-ary polymorphism that satisfies a certain term identity. Further, an arity of a polymorphism in the latter formulation was decreased to 4~\cite{4ary}. We will use the last fact as a key ingredient for our results. Very recently, several independent proofs of the Bulatov-Jeavons-Krokhin formulation were announced \cite{DBLP:Rafiey,Bulatov17a,Zhuk17}. Since the papers have not yet been thoroughly verified and widely accepted by the CSP community, in this paper we refer to the formulation as a hypothesis.

{\bf Related work.} A meta-problem of the VCSP topic is to establish the complexity of VCSP given that an input is restricted to an arbitrary subset of all input pairs $\left(\bR,\bGamma\right)$. A natural approach to this problem is to construct a new structure for any input $\left(\bR,\bGamma\right)$, $G_{\bR,\bGamma}$, and shift the analysis to $G_{\bR,\bGamma}$. In case of binary CSPs (i.e.\ when all relations of an input are binary) it is natural to define $G_{\bR,\bGamma}$ as a microstructure graph \cite{Jegou93a} of a template $\left(\bR,\bGamma\right)$. Thereby, a set of inputs, in which certain local substructures in $G_{\bR,\bGamma}$ are forbidden, forms a parametrized problem. Cooper and {\v{Z}}ivn\'y \cite{cz11:ai} investigated this formulation and found examples of specific forbidden substructures that result in tractable hybrid CSPs. Microstructure graphs also naturally appear in the context of fixed template CSPs. Specifically, if a template $\bGamma$ with binary relations is such that the arc and path consistency preprocessing of an instance of $\CSP\bGamma$ always results in a perfect microstructure graph, then additionally to satisfying all constraints (by finding a maximum clique) one can also optimize arbitrary sums of unary terms over a set of solutions (by assigning weights to vertices of the microstructure graph). The latter optimization problem is called {\em the minimum cost homomorphism problem} and all such templates were completely classified in \cite{Takhanov10adichotomy}.

Recently, a hybrid framework for VCSP has attracted some attention \cite{kolmogorov15:mrt}, that is when left structures are restricted to some set $\mathcal{H}$ and a right structure $\bGamma$ is fixed (the corresponding CSP is denoted as $\cspo_{\mathcal{H}}( \bGamma )$) and $\mathcal{H}$ is closed under inverse homomorphisms. The specific feature of this case is that for any input $\bR\in \mathcal{H}$ one can construct a new language $\bGamma_{\bR}$, called  
\ifTR
a lifted language (see Sec.~\ref{sec:construction}),
\else
a lifted language,
\fi
so that tractability of this language is a necessary condition for the tractability of $\cspo_{\mathcal{H}}( \bGamma )$.

{\bf Our results.} The first question that we address is a characterization of those templates $\bGamma$ for which the tractability of $\bGamma_{\bR}$ for any $\bR\in \mathcal{H}$ is not only necessary, but also is sufficient for the tractability of $\cspo_{\mathcal{H}}( \bGamma) $. We call $\bGamma$ that possesses this property for any $\mathcal{H}$ (closed under inverse homomorphisms) {\em widely tractable}. It turns out that the statement that the core of $\bGamma_{\bR}$ is preserved by a Siggers polymorphism (i.e. satisfies the Bulatov-Jeavons-Krokhin test for non-NP-hardness) is equivalent to the statement that $\bR$ is homomorphic to a certain structure $\bGamma'$ (constructed from $\bGamma$). Based on this observation we prove that, for a class of templates (that is likely to capture all templates), wide tractability is equivalent to the tractability of $\cspo_{\mathcal{H}_0}( \bGamma) $, where $\mathcal{H}_0$ is equal to a set of structures homomorphic to $\bGamma'$. Moreover, we prove that $\textsc{CSP}(\bGamma)$ can be in polynomial-time Turing reduced to $\textsc{CSP}(\bGamma')$ and, therefore, $\bGamma'$ is at least as hard as $\bGamma$. We develop an analogous theory for conservative valued CSPs.


\ifTR
{\bf Organization.} In Sec.~\ref{sec:prelim} we give all preliminary definitions and state theorems that we need. In Sec. 3 we describe an important construction called {\em a ``lifted language''}, taken from \cite{kolmogorov15:mrt}.
In subsection 4.1 we introduce the notion of widely tractable constraint language and in subsection 4.2 we prove necessary and sufficient conditions for wide tractability. A formulation and a proof of those conditions are based on the construction of a template $\bGamma'$ that we build from an initial fixed template $\bGamma$. We discuss properties of $\bGamma'$ in subsection 4.3. An analogous theory for conservative constraint languages, based on the corresponding construction of $\bGamma'_c$, is built in subsection 4.4.
\fi
\section{Preliminaries}\label{sec:prelim}
Throughout the paper it is assumed that $P \neq NP$. A problem is called {\em tractable} if it can be solved in polynomial time.
Let $\Qc=\mathbb{Q}\cup\{\infty\}$ denote the set
of rational numbers with (positive) infinity and $[k]=\{1,...,k\}$. Also, $D$ and $V$ are finite sets, $D^V$ is a set of mappings from $V$ to $D$.
We denote the tuples in lowercase boldface such as $\mathbf{a} = (a_1, \dots, a_k)$. Also for mappings $h \colon A \to B$ and tuples $\mathbf{a} = (a_1, \dots, a_k)$, where $a_j \in A$ for $j = 1, \dots, k$, we will write $\mathbf{b} = (h(a_1), \dots, h(a_k))$ simply as $\mathbf{b} = h(\mathbf{a})$. Relational structures are denoted in uppercase boldface as $\mathbf{R} = (R, r_1, \dots, r_k)$.
Finally let $\ar(\varrho)$, $\ar(\mathbf{a})$, and $\ar(f)$ stand for the arity of a relation $\varrho$, the size of a tuple $\mathbf{a}$, and the arity  of a function $f$, respectively.
\subsection{Fixed template VCSPs}
Let us formulate the general CSP as a homomorphism problem.

\begin{definition} Let $\mathbf{R} = (R, r_1, \dots, r_k)$ and $\mathbf{R}' = (R', r'_1, \dots, r'_k)$ be relational structures with a common signature (that is $\ar(r_i)= \ar(r'_i)$  for every $i = 1, \dots, k$). A mapping $h\colon R \to R'$ is called a \emph{homomorphism} from $\mathbf{R}$ to $\mathbf{R}'$ if for every $i = 1, \dots, k$ and for any $(x_1, \dots, x_{\ar(r_i)}) \in r_i$ we have that $\big((h(x_1), \dots, h(x_{\ar(r'_i)})\big) \in r'_i$. In that case, we write $\mathbf{R} \stackrel{h}{\to} \mathbf{R}'$ or sometimes just $\mathbf{R} \to \mathbf{R}'$.
\end{definition}

\begin{definition}{\bf The general CSP} is the following problem. Given a pair of relational structures with a common signature $\mathbf{R} = (V,r_1, \dots, r_k)$ and $\bGamma = (D, \varrho_1, \dots, \varrho_k)$, the question is whether there is a homomorphism $h:\mathbf{R} \to \bGamma$. The second structure $\mathbf{\Gamma}$ is called a template. 
\end{definition}

\begin{definition} Let $D$ be a finite set and $\bGamma$ be a finite relational structure over $D$. Then the {\bf fixed template CSP} for template $\bGamma$, denoted $\CSP\bGamma$, is defined as follows:
given a relational structure $\mathbf{R} = (V,r_1, \dots, r_k)$ of the same signature as $\bGamma$, the question is whether there is a homorphism $h:\mathbf{R} \to \mathbf{\Gamma}$. 
\end{definition}


A more general framework operates with {\em cost functions} $f:D^n \rightarrow \Qc$ instead of relations $\varrho\subseteq D^n$. 

\begin{definition}
Let us denote the set of all functions $f:D^n \rightarrow \Qc$ by $\CostF_D^{(n)}$ and let $\CostF_D=\bigcup_{n\ge 1}{\CostF_D^{(n)}}$.
We call the functions in $\CostF_D$ {\em cost functions} over $D$.
For every cost function $f\in\CostF_D^{(n)}$, let $\dom f=\{x\mid f(x)<\infty\}$. 
\end{definition}

\begin{definition}
An instance of the {\bf valued constraint satisfaction problem} (VCSP)
is a triple $(\bR, \bGamma, \{w_i(\bv)\}_{i\in [k], \bv\in r_i})$ where $\bR =  (V,r_1, \dots, r_k)$ is a relational structure, $\bGamma=(D,f_1,\ldots,f_k)$ is a tuple where $D$ is finite and $f_i\in \CostF_D^{(\ar(r_i))}$, $\{w_i(\bv)\}_{i\in [k], \bv\in r_i}$ are positive rationals, and the goal is to find an {\em assignment}  $h\in D^V$ that minimizes a function from $D^V$ to $\Qc$ given by
\begin{equation}\label{eq:VCSPinst}
f_{\calI}(h)=\sum_{i=1}^k\sum_{\bv\in r_i}{w_i(\bv)f_i(h(\bv))},
\end{equation}
A tuple $\bGamma=(D,f_1,\ldots,f_k)$ is called a \emph{valued template}.
\end{definition}

\begin{definition}
We will denote by \vcsp{\bGamma} a class of all VCSP instances in which the
valued template is $\bGamma$.
\end{definition}
For such $\bGamma$ we will denote by $\Gamma$ (without boldface) the set of cost functions $\{f_1,\ldots,f_k\}$.  A set $\Gamma$ is called a constraint language. 
The complexity of \vcsp{\bGamma} does not depend on the order of cost functions, therefore, we will use \vcsp{\Gamma} and \vcsp{\bGamma} interchangeably.

This framework captures many specific well-known problems,
including {\sc $k$-Sat}, {\sc Graph $k$-Colouring}, {\sc Minimum Cost Homomorphism Problem} and others (see~\cite{jeavons14:beatcs}).

A function $f\in\CostF_D^{(n)}$ that takes values in $\{0,\infty\}$ is called {\em crisp}. We will often view it as a relation in $D^n$,
and vice versa
(this should be clear from the context). If a language $\Gamma$ is crisp (i.e.\ it contains only crisp functions)
then $\VCSP\Gamma$ is a search problem corresponding to $\CSP\Gamma$. 

\begin{remark} Note that we formulated CSP as a decision problem, whereas VCSP as a search optimizational problem. This convention is followed throughout the text and further it becomes more important because decision and search problems are not computationally equivalent for hybrid CSPs (see after definition~\ref{4and3}).
\end{remark}

\begin{definition}
A constraint language $\Gamma$ (or, a template $\bGamma$) is said to be tractable, if
$\textsc{VCSP}(\Gamma_0)$ is tractable for each finite $\Gamma_0\subseteq\Gamma$. Also, $\Gamma$ (or, $\bGamma$) is NP-hard if there is a 
finite $\Gamma_0\subseteq\Gamma$ such that $\textsc{VCSP}(\Gamma_0)$ is NP-hard.
\end{definition}

An important problem in the CSP research is to characterize all tractable languages.
\subsection{Polymorphisms and fractional polymorphisms}

Let $\calO_D^{(m)}$ denote a set of all operations $g:D^m\rightarrow D$ and let
 $\calO_D=\bigcup_{m\ge 1}{\calO_D^{(m)}}$. 

Any language $\Gamma$ over a domain $D$ can be associated
with a set of operations on $D$,
known as the polymorphisms of $\Gamma$, defined as follows.
\begin{definition}
\label{def:polymorphism}
An operation $g\in \calO_D^{(m)}$ is a \emph{polymorphism} of a relation $\rho\subseteq D^n$ (or, $g$ \emph{preserves} $\rho$) if,
for any $\mathbf{x}^1,\ldots,\mathbf{x}^m \in \rho$,
we have that $g(\mathbf{x}^1,\ldots,\mathbf{x}^m)\in \rho$ where $g$ is applied component-wise.
For any crisp constraint language $\Gamma$ over a set $D$,
we denote by $\pol(\Gamma)$ a set of all operations on $D$ which are polymorphisms of every
$\rho \in \Gamma$.
\end{definition}
%

%
%
Polymorphisms play a key role in the algebraic approach to the CSP, but, for VCSPs, more general constructs are necessary, which we now define.

\begin{definition}
An $m$-ary \emph{fractional operation} $\omega$ on $D$ is a probability distribution on $\calO_D^{(m)}$.
The support of $\omega$ is defined as $\supp(\omega)=\{g\in \calO_D^{(m)}\mid \omega(g)>0\}$.
\end{definition}

\begin{definition}\label{def:fpol}
An $m$-ary fractional operation $\omega$ on $D$ is said to be a \emph{fractional polymorphism} of a
cost function $f\in \CostF_D$ if, for any $\mathbf{x}^1,\ldots,\mathbf{x}^m \in \dom f$,
we have
\begin{equation}
\sum_{g\in\supp(\omega)}{\omega(g)f(g(\mathbf{x}^1,\ldots,\mathbf{x}^m))} \le \frac{1}{m}(f(\mathbf{x}^1)+\ldots+f(\mathbf{x}^m)).
\label{eq:wpol-dist}
\end{equation}

For a constraint language $\Gamma$, $\fPol\Gamma$ will denote a set of all fractional operations that are fractional polymorphisms
of each function in $\Gamma$.
\end{definition}

We will also use symbols $\pol(\bGamma)$, $\fPol\bGamma$ meaning $\pol(\Gamma)$, $\fPol\Gamma$ respectively.
\subsection{Algebraic dichotomy conjecture}

An algebraic characterization for tractable templates was first conjectured by Bulatov, Krokhin and Jeavons~\cite{bulatov05:classifying},
and a number of equivalent formulations were later given in~\cite{MarotiMcKenzie,BartoKozikLics10,Siggers,4ary}.
We will use the formulation from~\cite{4ary} that followed a discovery by M. Siggers \cite{Siggers}; it is crucial for our purposes
that in the next definition an operation has a fixed arity (namely, 4) and, therefore, there is only a finite number of them on a finite domain $D$.

\begin{definition} An operation $s\colon D^4 \to D$ is called {\bf a Siggers operation on $D'\subseteq D$} if $s(x,y,z,t)\in D'$ whenever $x,y,z,t\in D'$ and for each $x,y,z \in D'$ we have:
\end{definition}

$$
\begin{array}{*{20}{c}}
{s(x,y,x,z) = s(y,x,z,y)}\\
{s(x,x,x,x) = x{\,\,\,\,\,\,\,\,\,\,\,\,\,\,\,\,\,\,\,\,\,\,\,}}
\end{array}
$$

\begin{definition} Let $g$ be a unary and $s$ be a 4-ary operations on $D$ and $g(D)=\left\{g(x)|x\in D\right\}$. A pair $(g,s)$ is called \emph{a Siggers pair on} $D$ if $s$ is a Siggers operation on $g(D)$.
A crisp constraint language $\Gamma$ is said to admit a Siggers pair $(g,s)$  if $g$ and $s$ are polymorphisms of $\Gamma$.
\end{definition}

\begin{theorem}[\cite{4ary}]\label{noSiggershard} 
A crisp constraint language $\Gamma$ that does not admit a Siggers pair is NP-Hard.
\end{theorem}


\begin{definition} A crisp language $\Gamma$ is called a \emph{BJK} language if it satisfies one of the following:
\begin{itemize}
\item $CSP(\Gamma)$ is tractable
\item $\Gamma$ does not admit a Siggers pair.
\end{itemize}
\end{definition}

{\bf Algebraic dichotomy conjecture}: Every crisp language $\Gamma$ is a BJK language.

This theorem first has been verified for domains of size 2 \cite{schaefer78:complexity}, 3 \cite{bulatov06:3-elementjacm}, or for languages containing all unary relations on $D$ \cite{bulatov11:conservative}. It has also been shown that it is equivalent to its restriction for directed graphs (that is when $\Gamma$ contains a single binary relation $\varrho$) \cite{DBLP:conf/cp/BulinDJN13}. Just recently, a number of authors  \cite{DBLP:Rafiey,Bulatov17a,Zhuk17} independently claimed the proof of the conjecture.

\section{Hybrid VCSP setting}\label{sec:construction}
\begin{definition} Let us call a family $\mathcal{H}$ of relational structures with a common signature 
 a {\bf structural restriction}.
\end{definition}

\begin{definition}[Hybrid CSP] Let $D$ be a finite domain, $\bGamma$ a template over $D$, and $\mathcal{H}$ a structural restriction of the same signature as $\bGamma$. We define $\cspo_\mathcal{H}(\bGamma)$ as the following problem: given a relational structure $\mathbf{R} \in \mathcal{H}$ as input, decide whether there is a homomorphism $h:\mathbf{R} \to \mathbf{\Gamma}$.
\end{definition}

\begin{definition}[Hybrid VCSP]\label{HybCSP} Let $D$ be a finite domain, $\bGamma = (D,f_1, \dots, f_k)$ a valued template over $D$, and $\mathcal{H}$ a structural restriction of the same signature as $\bGamma$.  We define $\vcspo_\mathcal{H}(\bGamma)$ as a class of instances of the following form.

An instance is a function from $D^V$ to $\Qc$ given by 
\begin{equation}\label{eq:VCSPinst}
f_{\calI}(h)=\sum_{i=1}^k\sum_{\bv\in r_i}{w_i(\bv)f_i(h(\bv))},
\end{equation}
where $\bR =  (V,r_1, \dots, r_k)\in\calH$ is a relational structure, $\{w_i(\bv)\}_{i\in [k], \bv\in r_i}$ are positive rationals.
The goal is to find an {\em assignment} $h\in D^V$ that minimizes $f_\calI$.
\label{def:hybridVCSPinst}
\end{definition}

The latter definition is too broad. Nonetheless, for certain classes of structural restrictions the tractability/intractability can be explained by algebraic means, and of special interest is the case when $\mathcal{H}$ is {\em up-closed}. 

\begin{definition} A family of relational structures $\mathcal{H}$ is called {\bf closed under inverse homomorphisms} (or {\bf up-closed} for short) if whenever $\mathbf{R}' \to \mathbf{R}$ and $\mathbf{R} \in \mathcal{H}$, then also $\mathbf{R}' \in \mathcal{H}$.
\end{definition}

Examples of hybrid CSPs with up-closed structural restrictions include such studied problems as a digraph $H$-coloring for an acyclic input digraph~\cite{swarts} or for an input digraph with odd girth at least $k$~\cite{kolmogorov15:mrt}, renamable Horn Boolean CSPs~\cite{GREEN20081094} and etc.
The key tool in their analysis is a construction of the so called lifted language that appeared first in~\cite{kolmogorov15:mrt}. In this construction, given arbitrary $\bR\in\calH$ one constructs a language $\Gamma_\bR$ over a finite domain, such that for tractability of $\textsc{VCSP}_\mathcal{H}(\bGamma)$, the tractability of $\textsc{VCSP}(\Gamma_\bR)$ is {\em necessary}.

Let us give a detailed description of $\Gamma_\bR$.
Given $\bR=(V,r_1,\ldots,r_k)$ and $\bGamma = (D, f_1,...,f_k)$ we define
$
D_\bR=V\times D
$ and $D_v = \left\{(v,a)|a\in D\right\}, v\in V$.


For tuples $\mathbf{a} = (a_1, \dots, a_p) \in D^p$ and $\bv=(v_1,\ldots,v_p)\in V^p$ denote $d(\mathbf{v}, \mathbf{a}) = ((v_1, a_1), ...,$ $(v_p, a_p))$.

Now for a cost function $f \in \CostF_D$ and $\mathbf{v} \in V^{\ar(f)}$ we will define a cost function on $D_\bR$ of the same arity as $f$ via
\begin{equation}\label{cons:pred}f^{\bv}(\mathbf{x})  = \begin{cases}
               f(\mathbf{y}) \quad \text{if} \,\, \mathbf{x} = d(\mathbf{v}, \mathbf{y}) \,\, \text{for some}\,\, \mathbf{y} \in D^{\ar(f)} \\
               \infty \qquad \text{otherwise} \\
            \end{cases}
\qquad\forall \mathbf{x}\in D_\bR^{\ar(f)}
\end{equation}
Finally, we construct the sought language $\Gamma_\bR$ on domain $D_\bR$ as follows:
$$\Gamma_\bR = \{f_i^\bv \::\: i\in[k], \bv \in r_i \} \cup \{ D_v \::\: v \in V\} 
$$
where relation
$D_v\subseteq D_\bR$ is treated as a unary function $D_v:D_\bR\rightarrow\{0,\infty\}$.

After ordering of its relations $\Gamma_{\bR}$ becomes a template $\bGamma_{\bR}$. 
The following is true~\cite{kolmogorov15:mrt}:
\begin{theorem}\label{vcspinherittract} Suppose that $\mathcal{H}$ is up-closed, $\bR \in \calH$ and $\bGamma$ is a (valued) template.
Then there is a polynomial-time reduction from $\textsc{(V)CSP}(\bGamma_\bR)$ to $\textsc{(V)CSP}_\mathcal{H}(\bGamma)$.
Consequently, \\
(a) 
if $\textsc{(V)CSP}_\mathcal{H}(\bGamma)$ is tractable then so is $\textsc{(V)CSP}(\bGamma_\bR)$; \\
(b) if $\textsc{(V)CSP}(\bGamma_\bR)$ is NP-hard then so is $\textsc{(V)CSP}_\mathcal{H}(\bGamma)$.
\end{theorem}

Let us give a proof of the latter theorem that slightly differs from the original one. 
For this purpose we will need a special case of hybrid VCSP, called {\em the VCSP with input prototype}. Given a finite relational structure $\bR$, denote $\up\left(\bR\right) = \left\{\bI|\bI\to \bR\right\}$.
\begin{definition}\label{4and3}
For a given valued template $\bGamma$ and a relational structure $\bR$ a problem $\textsc{VCSP}_{\mathcal{H}}(\bGamma)$ where $\mathcal{H} = \up\left(\bR\right)$ is called {\bf the VCSP with input prototype} $\bR$ and is denoted as $\textsc{VCSP}_{\bR}(\bGamma)$. If $\bGamma$ is crisp, then the decision version of $\textsc{VCSP}_{\bR}(\bGamma)$ is denoted as $\textsc{CSP}_{\bR}(\bGamma)$.
\end{definition}

It is easy to see that $\mathcal{H} = \up\left(\bR\right)$ is up-closed. Note that an input of $\textsc{(V)CSP}_{\bR}(\bGamma)$ is a relational structure $\bI$ that is homomorphic to $\bR$ but this homomorphism itself is not a part of the input. 
If we also assume that together with a structure $\bI$ we are given a homomorphism $h: \bI \to \bR$, then the latter problem is denoted as $\textsc{(V)CSP}^+_{\bR}(\bGamma)$.

\begin{remark} Note that the complexities of $\textsc{VCSP}_{\bR}(\bGamma)$ and $\textsc{VCSP}^+_{\bR}(\bGamma)$ can be sharply different. For example, consider 
$\bGamma = \left([4]; {\bf neq}_{4}\right)$ and $\bR = \left([3]; {\bf neq}_{3}\right)$ where ${\bf neq}_{k} = \{(i,j)| i,j\in[k], i\ne j\}$. While $\textsc{VCSP}_{\bR}(\bGamma)$, a problem of 4-coloring of a 3-colorable graph, is known to be NP-hard \cite{sanjeev}, $\textsc{VCSP}^+_{\bR}(\bGamma)$ is a trivial one. This example also demonstrates the distinction between decision and search in the hybrid framework: the decision problem $\textsc{CSP}_{\bR}(\bGamma)$ is also trivial, whereas its search version is NP-hard.
\end{remark}

\begin{lemma}\label{prototype} $\textsc{(V)CSP}(\bGamma_\bR)$ is polynomially equivalent to $\textsc{(V)CSP}^+_{\bR}(\bGamma)$
\end{lemma}
\ifTR
\begin{proof} {\bf Reduction of $\textsc{VCSP}(\bGamma_{\bR})$ to $\textsc{VCSP}^+_{\bR}(\bGamma)$.}
Let $\bGamma = \left(D, f_1,..., f_k\right)$ and $\bR=(V,r_1,\ldots,r_k)$ be given. 
An instance of $\textsc{VCSP}(\bGamma_{\bR})$ is a function:
$$\sum_{i\in[k], \bv \in r_i}\,\,\sum_{\bv'\in \rho_i^\bv}w_i^\bv(\bv')f_i^\bv(\bv')$$
where $\bI = (W, \langle\rho_i^\bv\rangle_{i\in[k], \bv \in r_i} )$ is an input structure whose $\rho_i^\bv$ corresponds to $f_i^\bv$ of $\bGamma_{\bR}$, and $\{w_i^{\bv} (\bv') | i\in[k], \bv \in r_i, \bv'\in \rho_i^\bv\}$ are positive rationals. 

Let us make the following consistency checking of that instance: we will check that for any variable $v\in W$ that is shared in two distinct terms $w_{i_1}^{\bv_1}(\bv'_1) f_{i_1}^{\bv_1}(\bv'_1)$ and $w_{i_2}^{\bv_2}(\bv'_2) f_{i_2}^{\bv_2}(\bv'_2)$ of the latter function  whether the projections of $\dom f_{i_1}^{\bv_1}$ and $\dom f_{i_2}^{\bv_2}$  on that variable have non-empty intersection. If they do not, we conclude that VCSP does not have solutions.

After that consistency checking, for our instance we can assume that there is an assignment $\delta: W\rightarrow V$, that assigns each variable $v\in W$ its domain $D_{\delta(v)}$. Denote ${\tilde\bI} = (W, \rho_1,...,\rho_k)$, where $\rho_i = \cup_{\bv\in r_i}\rho_i^\bv$.
It is easy to see that for any $\bv'\in \rho_i^\bv$ its component-wise image $\delta(\bv')$ is exactly the tuple $\bv$. Since $\bv\in r_i$, we conclude ${\tilde\bI}\mathop\rightarrow\limits^{\delta} \bR$. 

For $h: W\to D$, let us define $h^\delta: W \to V\times D$ by $h^\delta(v) = (\delta(v), h(v))$.
Vica versa,  to every assignment $h: W \to V\times D$ we will correspond an assignment $h^f(x) = F(h(x))$ where $F$ is a ``forgetting'' function, i.e. $F((v,a)) = a$.  For any assignment $h: W \to V\times D$ that satisfies $h(v)\in D_{\delta(v)}$, by construction, we have $(h^f)^\delta=h$. 
The expression to be minimized is
$$f_{\calI}(h) = \sum_{i\in[k]}\,\,\,\,\,\sum_{\bv \in r_i, \bv'\in \rho_i^\bv}w_i^\bv(\bv')f_i^\bv(\bv')$$
It is easy to see that if $h$ is an optimal solution of the latter sum, then $h^f$ is an optimal solution of the following
$$\sum_{i\in[k]}\sum_{\bv'\in \rho_i}(\sum_{\bv \in r_i}w_i^\bv(\bv'))f_i(\bv')$$
The latter is an instance of $\textsc{VCSP}_{\bR}(\bGamma)$ with input structure ${\tilde\bI}$ and a homomorphism $\delta:{\tilde\bI}\to \bR$, and the solution $s$ of it gives a solution $s^\delta$ of the initial one. Thus, we proved that $\textsc{VCSP}(\bGamma_{\bR})$ can be polynomially reduced to $\textsc{VCSP}^+_{\bR}(\bGamma)$.

{\bf Reduction of $\textsc{VCSP}^+_{\bR}(\bGamma)$ to $\textsc{VCSP}(\bGamma_{\bR})$.} Again, $\bGamma = \left(D, f_1,..., f_k\right)$, $\bR=(V,r_1,\ldots,r_k)$. Suppose we are given an instance of $\textsc{VCSP}^+_{\bR}(\bGamma)$ with an input structure $\bI = (W, \rho_1,...,\rho_k)$ and a homomorphism $\delta:\bI\to \bR$, i.e. our goal is to minimize:
$$f_{\calI}(h)=\sum_{i=1}^k\sum_{\bv\in \rho_i}{w_i(\bv)f_i(h(\bv))}$$
over $h\in D^W$.
We can construct an instance of $\textsc{VCSP}(\bGamma_{\bR})$:
$$f_{\calI}(s)=\sum_{i=1}^k\sum_{\bv\in \rho_i}{w_i(\bv)f^{\delta(\bv)}_i(s(\bv))}$$
where $s: W\to V\times D$ is such that $s(v)\in D_{\delta(v)}$ (these constraints can be modeled via crisp functions $D_i\in \Gamma_{\bR}$). It is straightforward to check that if $s$ is a solution of $\textsc{VCSP}(\bGamma_{\bR})$ then $h = s^f$ is a solution of $\textsc{VCSP}_{\bR}(\bGamma)$.

A proof of the equivalence of $\textsc{CSP}(\bGamma_\bR)$ and $\textsc{CSP}^+_{\bR}(\bGamma)$ for crisp $\bGamma$
can be done analogously.
\end{proof}
\else
\fi

\begin{proof}[Theorem~\ref{vcspinherittract} (a)] Since $\mathcal{H}$ is up-closed, then for any $\bR\in \mathcal{H}$, $\{\bI|\bI\to\bR\}\subseteq\mathcal{H}$. I.e. a problem $\textsc{VCSP}_{\bR}(\bGamma)$ is a restriction of $\textsc{VCSP}_{\mathcal{H}}(\bGamma)$ to certain inputs. Therefore, $\textsc{VCSP}^+_{\bR}(\bGamma)$ is polynomially reducible to $\textsc{VCSP}_{\mathcal{H}}(\bGamma)$. Using the previous lemma, we conclude that for the tractability of $\textsc{VCSP}_{\mathcal{H}}(\bGamma)$ it is necessary that $\textsc{VCSP}^+_{\bR}(\bGamma)$ and $\textsc{VCSP}(\bGamma_{\bR})$ are tractable. Part (b) can be proved analogously.
\end{proof}



\section{Wide tractability of a crisp language}\label{sec:result}
Throughout this section we will assume that $\bGamma$ is crisp. 
\subsection{Widely tractable languages}
For up-closed structural restrictions $\mathcal{H}$, the construction of a lifted language gives us the necessary conditions for the tractability of $\textsc{CSP}_\mathcal{H}(\bGamma)$ (Theorem \ref{vcspinherittract} (a)). Let us now define {\em widely tractable templates} $\bGamma$ as those for which the necessary conditions for the tractability of $\textsc{CSP}_\mathcal{H}(\bGamma)$ are, in fact, sufficient:

\begin{definition} A template $\bGamma$ is called {\bf widely tractable} if for any up-closed $\mathcal{H}$, $\textsc{CSP}_\mathcal{H}(\bGamma)$ is tractable if and only if $\textsc{CSP}(\bGamma_\bR)$ is tractable for any $\bR\in \mathcal{H}$.  
\end{definition}

The concept of wide tractability is important in the hybrid CSPs setting due to the following theorem:

\begin{theorem}\label{maximality} If a  template $\bGamma$ is widely tractable, then there is an up-closed $\mathcal{H}^{\bGamma}$ such that for any up-closed $\mathcal{H}$, $\textsc{CSP}_\mathcal{H}(\bGamma)$ is tractable if and only if $\mathcal{H}\subseteq \mathcal{H}^{\bGamma}$.
\end{theorem}

\begin{proof}
Let us define
\begin{equation}\label{str:maximal}
\mathcal{H}^{\bGamma} = \left\{\bR | \textsc{CSP}(\bGamma_{\bR})\textit{ is tractable}\right\}
\end{equation}
It is easy to see that $\mathcal{H}^{\bGamma}$ is up-closed itself.
By definition, $\mathcal{H}^{\bGamma}$ contains only such $\bR$ for which $\textsc{CSP}(\bGamma_{\bR})$ is tractable, and this together with wide tractability of $\bGamma$, implies that $\textsc{CSP}_{\mathcal{H}^{\bGamma}}(\bGamma)$ is tractable.

Suppose that for some up-closed $\mathcal{H}$, $\textsc{CSP}_\mathcal{H}(\bGamma)$ is tractable. From the wide tractability of $\bGamma$ we obtain that it is equivalent to stating that $\textsc{CSP}(\bGamma_\bR)$ is tractable for any $\bR\in \mathcal{H}$. But the last is equivalent to  $\mathcal{H}\subseteq \mathcal{H}^{\bGamma}$.
\end{proof}

\subsection{Wide tractability in case of strongly BJK languages}

In this section we will give necessary and sufficient conditions of wide tractability in a very important case of crisp languages, namely, {\em strongly BJK} languages. 

\begin{definition} A crisp language $\bGamma$ is called strongly BJK language if for any $\bR$ the lifted $\bGamma_{\bR}$ is BJK.
\end{definition}
\begin{remark} As we have already noted it is likely that this class includes all crisp languages~\cite{DBLP:Rafiey,Bulatov17a,Zhuk17}. 
\end{remark}

\ifTR
\else
\fi





Before introducing the main theorem of this section, let us describe one construction. Let $\rho$ be some $m$-ary relation over a domain $D$. It induces a new relation $\rho'$ over a set of Siggers pairs on a set $D$, denoted $D'$, by the following rule: a tuple of Siggers pairs $\big((g_1, s_1), \cdots, (g_m, s_m)\big)\in\rho'$ if and only if for any $\left(x_1,...,x_m\right)\in\rho$ we have that $\left(g_1(x_1), ..., g_m(x_m)\right)\in \rho$ and for any tuples  
$\left(a_1,...,a_m\right)$, $\left(b_1,...,b_m\right)$, $\left(c_1,...,c_m\right)$, $\left(d_1,...,d_m\right)$ from $\rho$ we have that $\big(s_1(a_1,b_1,c_1,d_1)$, ... , $s_m(a_m,b_m,c_m,d_m)\big)\in \rho$. 


Given a relational structure $\bGamma=\left(D,\rho_1,...,\rho_s\right)$, we define $\bGamma' = \left(D',\rho'_1,...,\rho'_s\right)$.

\begin{theorem}\label{theorem:main} Let $\bGamma$ be a strongly BJK language. Then $\bGamma$ is widely tractable if and only if  $\textsc{CSP}_{\bGamma'}(\bGamma)$ is tractable.
\end{theorem}
A proof of theorem~\ref{theorem:main} is mainly based on the following lemma:
\begin{lemma} \label{homtoG} For an arbitrary $\bR$, $\bGamma_\bR$ admits a Siggers pair if and only if there is a homomorphism $h:\bR\rightarrow \bGamma'$.
\end{lemma}
\ifTR
\begin{proof} Let $\bR=\big(V,r_1, ..., r_k\big)$. For $\bGamma_\bR$ that admits a Siggers pair $(g, s)$, let us construct a homomorphism $h:\bR\rightarrow\bGamma'$. 
Recall that $\bGamma_{\bR}$ is defined over a domain $D_\bR=\cup_{v\in V}D_v$ (see the definition of $D_v$ in subsection \ref{sec:construction}). Let us now define restrictions of $g$ and $s$ on $D_v$, i.e. $g_v=g|_{D_v}$ and $s_v=s|_{D_v}$ (this is possible because $g, s$ preserve $D_v$). In turn, $g_v$ and $s_v$ correspond to operations $g'_v$ and $s'_v$ defined on $D$ that satisfy $g'_v(a)=g_v\big((v,a)\big)$ and $s'_v(a,b,c,d)=s_v\big((v,a), (v,b), (v,c), (v,d)\big)$. Let us denote as $h$ a mapping $v\mapsto (g'_v, s'_v)$. It is easy to see that $h$ maps $V$ to $D'$,  the domain of $\bGamma'$. 

Let us show that $h$ is a homomorphism from $\bR$ into $\bGamma'$. Consider $\bv=(v_1,...,v_p)\in r_i$ and $\rho^\bv_i\in\Gamma_{\bR}$ (see the definition of $\rho^\bv_i$ in subsection \ref{sec:construction}). Since $g, s$ preserve $\bGamma_{\bR}$, we conclude that $g, s$ preserve $\rho^\bv_i$. I.e., for any $\left(x_1,...,x_p\right)\in \rho^\bv_i$ we have that $\left(g_{v_1}(x_1), ..., g_{v_p}(x_p)\right)\in \rho^\bv_i$ and for any  $\left(a_1,...,a_p\right)$, $\left(b_1,...,b_p\right)$, $\left(c_1,...,c_p\right)$, $\left(d_1,...,d_p\right)\in \rho^\bv_i$ we have that $\big(s_{v_1}(a_1,b_1,c_1,d_1)$, ... , $s_{v_p}(a_p,b_p,c_p,d_p)\big)\in \rho^\bv_i$. 
If we identify element $(v,a)$ of $D_{v}$ with element $a$ of $D$ (for all $v, a$), in the last condition we have to change $D_{v_i}$ to $D$ and $s_v$ to $s'_v$ and $\rho^\bv_i$ to $\rho_i$. I.e., the condition will become equivalent to the statement that $\big((g'_{v_1}, s'_{v_1}), \cdots, (g'_{v_p}, s'_{v_p})\big)$, i.e. $\big( h(v_1),...,h(v_p)\big)$, is in $\rho'_i$. We proved that for any $\bv=(v_1,...,v_p)\in r_i$ its image is in $\rho'_i$, i.e. $h$ is a homomorphism.

Thus, we proved that if $\bGamma_\bR$ admits a Siggers pair, then there is a homomorphism $h:\bR\rightarrow \bGamma'$. Suppose now that for some $\bR$, $o: \bR\rightarrow \bGamma'$ is a homomorphism. Let us define a Siggers pair $(g_o, s_o)$ on $D_{\bR}$ in such a way that $(g_o|_{D_v}, s_o|_{D_v})$ coincides with $o(v)$ if we identify $(v,a)\in D_v$ and $a\in D$. It is straightforward to check that  $\bGamma_{\bR}$ admits $(g_o, s_o)$.
\end{proof}

\begin{proof}[Proof of Theorem \ref{theorem:main}] Suppose that $\bGamma$ is widely tractable. 
Let us define
\begin{equation}\label{str:maximal}
\mathcal{H}^{\bGamma} = \left\{\bR | \textsc{CSP}(\bGamma_{\bR})\textit{ is tractable}\right\}
\end{equation}
Since $\bGamma$ is strongly BJK, we obtain that 
\begin{align*}
\mathcal{H}^{\bGamma} = \left\{\bR | \bGamma_{\bR}\textit{ admits a Siggers pair}\right\}
\end{align*}
Therefore, from lemma \ref{homtoG} we conclude that $\mathcal{H}^{\bGamma} = \up(\bGamma')$.

By definition, $\mathcal{H}^{\bGamma}$ contains only such $\bR$ for which $\textsc{CSP}(\bGamma_{\bR})$ is tractable, and this together with the wide tractability of $\bGamma$, implies that $\textsc{CSP}_{\mathcal{H}^{\bGamma}}(\bGamma)=\textsc{CSP}_{\bGamma'}(\bGamma)$ is tractable.

Suppose now that $\textsc{CSP}_{\bGamma'}(\bGamma)$ is tractable. Let us prove that $\bGamma$ is widely tractable, i.e. let us verify that from the tractability of $\textsc{CSP}(\bGamma_\bR)$ for any $\bR\in  \mathcal{H}$ we can deduce that $\textsc{CSP}_\mathcal{H}(\bGamma)$ is tractable.
Suppose that $\textsc{CSP}(\bGamma_\bR)$ is tractable for any $\bR\in\mathcal{H}$. Thus, due to the strong BJK property,  $\bGamma_\bR$ admits a Siggers pair. From lemma \ref{homtoG} we obtain that $\bR\in \up( \bGamma')$, i.e. $\mathcal{H}\subseteq \up(\bGamma')$, and $\textsc{CSP}_\mathcal{H}(\Gamma)$ is tractable.
\end{proof}
\else
\fi
\begin{remark} If $\bGamma'\to \bGamma$ then $\textsc{CSP}_{\bGamma'}(\bGamma)$ is a trivial problem. In the latter case theorem~\ref{theorem:main} gives us that $\bGamma$ is a widely tractable template. Such templates are quite common. E.g. our computational experiment showed (see section~\ref{exper}) that if $D=\{0,1\}$ and $\rho\subseteq \{0,1\}^3$ is such that $\Gamma=\{\rho\}$ is NP-hard, then $\bGamma'\to \bGamma$. Example of a widely tractable and NP-hard $\bGamma$ for which $\bGamma'\not\to \bGamma$ will be given in the next section (example~\ref{betweennes}).
\end{remark}

\subsection{Relationship between $\bGamma$ and $\bGamma'$}

The binary relation $\to$ is transitive, reflexive, but not antisymmetric. It also induces the equivalence relation $\sim$ on a set of all finite structures: 
$$\bR_1\sim \bR_2 \Leftrightarrow \bR_1\to \bR_2, \bR_2\to \bR_1$$

\begin{theorem}\label{below} For any $\bGamma$, $\bGamma\to \bGamma'$.
\end{theorem}
\ifTR
\begin{proof} For any $a\in D$, let $a'$ be a Siggers pair $(a,a)$ where the first element is understood as a unary constant operation and the second element as a 4-ary constant operation. Thus, $a'\in D'$, and we can define a function $h$ by equality 
$h(a)=a'$. Let us prove that $h$ is a homomorphism from $\bGamma$ to $\bGamma'$. 

For any $(d_1,...,d_p)\in\rho_i$ its image is $\big(d'_1, ... , d'_p\big)$. We need to check that the last tuple is in $\rho'_i$. Indeed, if we recall the definition of a certain tuple $\big((g_1, s_1), ... , (g_m, s_m)\big)$ being in $\rho'_i$, it can be reduced to the statement that of the kind: $(g_1(...),...,g_m(...))\in \rho_i$ (and $(s_1(...),...,s_m(...))\in \rho_i$). But in our case the latter conditions trivially hold.
\end{proof}
\else
\fi







Thus, we can view $\textsc{CSP}(\bGamma')$ as a relaxation of $\textsc{CSP}(\bGamma)$. Moreover, the theorem \ref{below} has the following interesting consequence.

\begin{theorem}\label{reduce}
If $\bGamma$ is strongly BJK, then there is a polynomial-time Turing reduction from $\textsc{CSP}(\bGamma)$ to $\textsc{CSP}(\bGamma')$
\end{theorem}
\ifTR
\begin{proof} From lemma~\ref{homtoG} we obtain that $\bGamma_{\bGamma'}$ admits a Siggers pair. Since $\bGamma$ is strongly BJK, then $\textsc{CSP}(\bGamma_{\bGamma'})$ is tractable. Lemma~\ref{prototype} gives that hybrid $\textsc{CSP}^+_{\bGamma'}(\bGamma)$ is also tractable.

Let us describe our reduction. Given an input $\bR$ for $\textsc{CSP}(\bGamma)$ we first check whether $\bR\in \up(\bGamma')$. If $\bR\notin \up(\bGamma')$ then due to theorem \ref{below} we can answer that $\bR\notin \up(\bGamma)$. Alternatively, if $\bR\in \up(\bGamma')$, we will be given a homomorphism $h:\bR\to \bGamma'$ (using that for fixed template CSPs search and decision problems are polynomially equivalent) and can reduce our problem to $\textsc{CSP}^+_{\bGamma'}(\bGamma)$. Therefore, we can identify in polynomial time whether $\bR\in \up(\bGamma)$ or not.
\end{proof}
\else
\fi

If $\bGamma$ is tractable, then $\bGamma'$ is preserved by a constant $(g,s)\in D'$, where $(g,s)$ is a Siggers pair that is admitted by $\bGamma$. I.e., $\up(\bGamma')$ is a set of all finite structures with the same vocabulary as $\bGamma$. We can take any tractable $\bGamma$ that is not constant-preserving (e.g. $\bGamma = \left([3]; {\bf neq}_3\right)$) as an example of a template for which $\bGamma \not\sim \bGamma'$, i.e. $\bGamma'\not\to\bGamma$. 

The following example demonstrates an NP-hard $\bGamma$ for which $\bGamma \not\sim \bGamma'$.

\begin{example}\label{betweennes} Define $\bGamma = \left(\left\{0,1\right\}; \left\{0\right\}, \left\{1\right\}, \rho\right)$, where $\rho = \left\{0,1\right\}^3\setminus \left\{(0,1,0),(1,0,1)\right\}$. A fixed-template $\textsc{CSP}$ with this $\bGamma$ is called \emph{the boolean betweennes}, and it is NP-hard because $\bGamma$ does not fall into any of Schaefer`s classes~\cite{schaefer78:complexity}. 

The boolean betweennes can be popularly reformulated in the following way. Suppose that we have a number of $n$ towns $v_1, ..., v_n$ and a system of roads (each consisting of 3 consecutive towns) $(v_{\alpha_1}, v_{\alpha_2}, v_{\alpha_3}), ..., (v_{\omega_1}, v_{\omega_2}, v_{\omega_3})$. Our goal is to divide those towns between 2 states (assign 0 or 1 to $n$ variables) in such a way that unary constraints are satisfied, i.e. certain towns should be given to prespecified states, and every road should not cross administrative barriers twice.

Let $\bGamma_{\alpha} = \left(\left\{0,1,\alpha\right\}; \left\{0,\alpha\right\}, \left\{1,\alpha\right\}, \rho_{\alpha}\right)$, where $\rho_{\alpha} = \rho\cup \{(1,1,\alpha),(\alpha,1,1),(0,0,\alpha), (\alpha, 0, 0), \\ (0, \alpha, 1), (1,\alpha, 0)\}$. 
A symbol $\alpha$ can be interpreted as a ``dual attachment'' status that can be given to towns, for which we can freely change $\alpha$-status to both 0 and 1 without violating ternary constraints.

It is easy to see that $\bGamma_{\alpha}\not\rightarrow \bGamma$ (image of $\alpha$ cannot be both 0 and 1). If we prove that $\textsc{CSP}(\bGamma_{\bGamma_{\alpha}})$ is tractable (and, therefore, $\bGamma_{\bGamma_{\alpha}}$ admits a Siggers pair), this will lead to a conclusion that $\bGamma_{\alpha}\rightarrow\bGamma'$ by lemma \ref{homtoG}, and consequently, $\bGamma'\not\rightarrow \bGamma$. 

According to lemma~\ref{prototype}, $\textsc{CSP}(\bGamma_{\bGamma_{\alpha}})$ is equivalent to a problem of deciding whether there is a homomorphism $h:\bR\rightarrow \bGamma$ for a relational structure $\bR = (V, \Omega_0, \Omega_1, \Omega)$ and a homomorphism $g:\bR\rightarrow \bGamma_{\alpha}$ given as inputs. If $\Omega_0\cap \Omega_1\ne\emptyset$ we claim the nonexistence of $h$. Otherwise, $h$ is defined in the following way: $h(x) = g(x)$, if $g(x)\ne \alpha$; $h(x) = 0$, if $x\in \Omega_0$ and $g(x)=\alpha$; $h(x) = 1$, if $x\in \Omega_1$ and $g(x)=\alpha$; and $h(x) = 0$, if otherwise. It can be checked that this algorithm solves $\textsc{CSP}^+_{\bGamma_{\alpha}}(\bGamma)$.

Our computational experiment showed (see section~\ref{exper}) showed that, in fact, $\bGamma'\sim \bGamma_{\alpha}$. It is easy to see that in the latter algorithm for $\textsc{CSP}(\bGamma_{\bGamma_{\alpha}})$ we used a homomorphism $g:\bR\rightarrow \bGamma_{\alpha}$ only at the stage of the construction of $h$, i.e. we did not need it at the decision stage. The latter means that $\textsc{CSP}_{\bGamma_{\alpha}}(\bGamma)$ as a decision problem is also tractable and from theorem~\ref{theorem:main} we obtain that $\bGamma$ is widely tractable (under condition that it is strongly BJK).
\end{example}

\ifTR
Let us also give an example of a class of languages $\bGamma$, for which $\bGamma'$ is provably NP-hard, without strong BJK assumptions on $\bGamma$. 

\begin{example} Let us consider $H$-coloring problem, i.e.  $\textsc{CSP}(\Gamma=\left\{H\right\})$, where $H$ is an irreflexive symmetric relation. A famous result of  Hell and Ne\v{s}et\v{r}il~\cite{Hell} states that $\textsc{CSP}(H)$ is tractable if $H$ is bipartite, and NP-hard, otherwise. Consider a case of non-bipartite $H$.
It is easy to see from the construction of $\bGamma'$ that it contains 1 symmetric relation $H'$. It also should be irreflexive, because, otherwise, $H$ would be preserved under some Siggers pair. Due to theorem \ref{below}, there is a homomorphism from $H$ to $H'$, therefore, $H'$ is also non-bipartite. Therefore, $\textsc{CSP}(H')$ is NP-hard.

An open problem is to find a language in this class for which $\bGamma \not\sim \bGamma'$. This problem is connected with a question whether there are graphs $H_1, H_2$ such that $H_2\not\to H_1$ and $H_1$-coloring is tractable in the class of $H_2$-colorable graphs. 
So far it is known that even 3-colorability of 4-colorable graph is NP-hard. Moreover, determining whether a graph is 3-colorable remains NP-hard for triangle-free graphs with maximum degree 4 \cite{Maffray} (the latter, by Brooks theorem, are 4-colourable).
\end{example} 
\else
\fi

Theorem \ref{reduce} gives us the idea that we can reduce $\textsc{CSP}(\bGamma)$ to $\textsc{CSP}(\bGamma')$, $\textsc{CSP}(\bGamma')$ to $\textsc{CSP}(\bGamma'')$ and etc. It turns out that this sequence of reductions collapses very soon:

\begin{theorem}\label{doublereduce}
If $\bGamma, \bGamma'$ are both strongly BJK, then $\bGamma'\sim\bGamma''$.
\end{theorem}
\ifTR
\begin{proof}
Since $\bGamma, \bGamma'$ are strongly BJK, lemmas \ref{prototype}-\ref{homtoG} give us that $\textsc{CSP}^+_{\bGamma'}(\bGamma)$ and $\textsc{CSP}^+_{\bGamma''}(\bGamma')$ are tractable.

Let us show that $\textsc{CSP}^+_{\bGamma''}(\bGamma)$ is tractable. Let $\bR$ and $h:\bR\to \bGamma''$ be an input to $\textsc{CSP}^+_{\bGamma''}(\bGamma)$. Since $\textsc{CSP}^+_{\bGamma''}(\bGamma')$ is tractable, we can check in polynomial time whether $\bR\in \up(\bGamma')$. If $\bR\notin \up(\bGamma')$, then from theorem \ref{below} we conclude that  $\bR\notin \up(\bGamma)$. Alternatively, if $\bR\stackrel{h'}{\to}\bGamma'$ (here we again use the polynomial equivalence if search and decision problems for fixed template CSPs), we can input $(\bR, h')$ to  $\textsc{CSP}^+_{\bGamma'}(\bGamma)$ and polynomially check whether $\bR\in \up(\bGamma)$.

Since  $\textsc{CSP}^+_{\bGamma''}(\bGamma)$ is tractable, then $\textsc{CSP}(\bGamma_{\bGamma''})$ is tractable and $\bGamma_{\bGamma''}$ admits a Sigger pair. From lemma \ref{homtoG} we conclude that $\bGamma''\to \bGamma'$, i.e. $\bGamma''\sim\bGamma'$.
\end{proof}
\else
\fi

\section{Valued templates: conservative case}

So far, the most applicable class of fixed-template valued VCSPs was {\em the submodular function minimization} problems \cite{Kolmogorov:2002}. Also, {\em minimum cost homomorphism problems (MinHom)} appeared in such different contexts as Defense Logistics \cite{LORA} and Computer Vision \cite{ECCV_Best_paper}. These two examples make the framework of {\em conservative valued CSPs} of special interest, since it includes both MinHom and submodular function minimization. 
The structure of tractable conservative languages is very clearly understood both in crisp~\cite{bulatov11:conservative} and valued cases~\cite{Thapper15:Sherali}. Let us now give the definition. 

\begin{definition} A valued constraint language $\Gamma$ is called \emph{conservative} if it contains  
${\bf Un}_D$, where  
${\bf Un}_D$ is a set of all unary $\{0,1\}$-valued cost functions over $D$.
\end{definition}

In the hybrid VCSPs setting, if the right structure $\bGamma$ is conservative, we have to make a certain supplementary assumption on structural restrictions, so that we do not loose the desirable property that optimized function can have an arbitrary unary part.

\begin{definition} We say that a relational structure $\mathcal{H}$ {\bf does not restrict unaries} if for each $\mathbf{R} \in \mathcal{H}$
of the form $\mathbf{R} = (V, r_1, \dots, r_{i-1}, r_i, r_{i+1}, \dots, r_{k})$ with $\ar(r_i)=1$ and for each unary relation $r'_i\subseteq V$, we have
 $\mathbf{R}' \in \mathcal{H}$, where $\mathbf{R}' = (V, r_1, \dots, r_{i-1}, r'_i, r_{i+1}, \dots, r_{k})$.
\end{definition}

A generalization of the wide tractability for conservative languages will be the following definition.

\begin{definition} A valued conservative language $\bGamma$ is called {\bf widely c-tractable} if for any up-closed $\mathcal{H}$ that does not restrict unaries, $\textsc{VCSP}_\mathcal{H}(\bGamma)$ is tractable if and only if $\textsc{VCSP}(\bGamma_\bR)$ is tractable for any $\bR\in \mathcal{H}$.  
\end{definition}

\ifTR
Recently, Thapper and \v{Z}ivn\'y \cite{Thapper15:Sherali} introduced a generalization of relational width for valued languages, which they called \emph{a valued relational width}. 
Their generalization is based on reducing VCSPs to the linear programming relaxations (the so called Sherali-Adams relaxation) parameterized by $(k, l), k,l\in {\mathbb N}, k < l$.
For completeness, let us formulate this relaxation. 
Any function from the definition \ref{HybCSP} can be represented in the following form:
$$\sum_{i=1}^q \phi_i(S_i)$$
where $S_i\subseteq V$ and $\phi_i:D^{|S_i|}\rightarrow \overline{Q}$. In the expression $\phi_i(S_i)$ we assume that variables from $S_i$ come in a certain order.
We assume that for any subset of variables $S$ such that $|S|\leq k$ 
there is a term $\phi(S)$ in our function. We can make this possibly by adding constant-0 weighted cost functions.

For any $i\in [q]$ and ${\bf s}\in D^{|S_i|}$ a variable $\lambda_i({\bf s})$ is introduced and the following expression
\begin{equation}
\sum_{i=1}^q{\lambda_i({\bf s}) \phi_i({\bf s})}
\end{equation}
is minimized under conditions:
$$\lambda_i({\bf s}) \geq 0, \forall i\in [q], {\bf s}\in D^{|S_i|}$$
$$\lambda_i({\bf s}) = 0, \forall i\in [q], {\bf s}\notin \dom (\phi_i)$$
$$\sum_{{\bf s}\in \dom(\phi_i)} \lambda_i({\bf s}) = 1$$
$$\lambda_j({\bf t}) = \sum_{(S_j,{\bf t}) \subseteq (S_i,{\bf s})}\lambda_i({\bf s}), \forall i, j\in [q]: S_j\subseteq S_i , \ar(\phi_j)\leq k, {\bf t}\in D^{\ar(\phi_j)}$$
where $(S_j,{\bf t}) \subseteq (S_i,{\bf s})$ means that an assignment of variables $S_j$ to $\bf t$ is consistent with an assignment of variables $S_i$ to $\bf s$.
Thapper and \v{Z}ivn\'y showed that the Sherali-Adams relaxation with parameters $(2,3)$ solves any tractable fixed-template valued conservative CSP. Moreover, the complexity of the latter algorithm depends on the size of a template $\bGamma$ polynomially. This implies the following result.
\else
\fi
\begin{theorem} Any conservative valued language is widely c-tractable.
\end{theorem}
\ifTR
Before we start the proof of this theorem, we need the lemma:
\begin{lemma} If $\bGamma$ is conservative and $\bR$ is such that $\textsc{VCSP}(\bGamma_{\bR})$ is tractable, then $\Gamma_{\bR}\cup {\bf Un}_{V\times D}$ is a tractable conservative language.
\end{lemma}
\begin{proof}[Sketch] In lemma \ref{prototype} we showed the equivalence of $\textsc{VCSP}(\bGamma_{\bR})$ and $\textsc{VCSP}^+_{\bR}(\bGamma)$. Therefore, $\textsc{VCSP}^+_{\bR}(\bGamma)$ is a tractable problem.

Now let us repeat all arguments from lemma \ref{prototype} step by step and apply them to an instance of $\textsc{VCSP}(\Gamma_{\bR}\cup {\bf Un}_{V\times D})$. 
It can be checked that the final expression (that is to be minimized) will contain terms with cost functions from $\Gamma$ plus a sum of arbitrary unary terms. But since $\Gamma$ is conservative, the problem is an instance of $\textsc{VCSP}^+_{\bR}(\bGamma)$. Thus, $\Gamma_{\bR}\cup {\bf Un}_{V\times D}$ is tractable.
\end{proof}

\begin{proof}[theorem] That is we have to prove that if $\bGamma$ is conservative and $\mathcal{H}$ is up-closed and does not restrict unaries, then either
\begin{itemize}
\item there is $\bR\in \mathcal{H}$ such that $\textsc{VCSP}(\bGamma_{\bR})$ is intractable, or
\item $\textsc{VCSP}_\mathcal{H}(\bGamma)$ is tractable.
\end{itemize}

It is sufficient to prove that $\textsc{VCSP}_\mathcal{H}(\bGamma)$ is tractable if for any 
$\bR\in \mathcal{H}, \bGamma_{\bR}$ is tractable.

Suppose we are given an instance $\bR\in \mathcal{H}$ of $\textsc{VCSP}_\mathcal{H}(\bGamma)$. Now we can simply input $\bR$ together with the identity homomorphism $h: \bR\to \bR$ to $\textsc{VCSP}^+_\bR(\bGamma)$ and find the optimal solution. 

The latter problem is equivalent to $\textsc{VCSP}(\bGamma_\bR)$ and is tractable. From lemma we obtain that $\Gamma_{\bR}\cup {\bf Un}_{V\times D}$ is conservative and tractable. Therefore, $\textsc{VCSP}(\bGamma_\bR)$ can be solved by Sherali-Adams relaxation in time polynomial from the size of $\bGamma_\bR$ and an input structure. It is easy to see that the total number of steps in this strategy depends on the size of $\bR$ polynomially. 
\end{proof}
\else
\fi

An analog of theorem \ref{maximality} is the following statement.

\begin{theorem} \label{max-cons} For any conservative valued language $\bGamma$ there is an up-closed $\mathcal{H}^{\bGamma}_{c}$ that does not restrict unaries and such that for any up-closed $\mathcal{H}$  that does not restrict unaries, $\textsc{VCSP}_\mathcal{H}(\bGamma)$ is tractable if and only if $\mathcal{H}\subseteq \mathcal{H}^{\bGamma}_{c}$.
\end{theorem}
\ifTR
\begin{proof}[sketch]
We will use wide c-tractability of any conservative valued language.
The statement can be proved by absolutely analogous arguments as theorem \ref{maximality}. We only have to note that a set
\begin{equation}\label{max-cons}
\mathcal{H}^{\bGamma}_{c}= \left\{\bR | \textsc{VCSP}(\bGamma_{\bR}\cup\bf{Un})\textit{ is tractable}\right\}
\end{equation}
does not restrict unaries.
\end{proof}
\else
\fi

Our next goal will be to prove that $\mathcal{H}^{\bGamma}_{c}=\up(\bGamma'_{c})$ for a certain template $\bGamma'_{c}$. 
If in a case of $\textsc{CSP}_\mathcal{H}(\bGamma)$ we used a description of tractable templates in terms of polymorphisms, in the current case we will need a description via fractional polymorphisms.

\begin{definition}
Let $(\sqcup,\sqcap)$ be a pair of binary operations and $(Mj_1, Mj_2, Mn_3)$ be a triple
of ternary operations defined on a domain $D$, and $M\subseteq \left\{\left\{a,b\right\}|a,b\in D,a\ne b\right\}$.

The pair $(\sqcup,\sqcap)$, is a {\bf symmetric tournament polymorphism (STP)} on $M$ if $\forall x,y$, $\{x\sqcup y, x\sqcap y\} = \{x,y\}$ and for any $\left\{a,b\right\}\in M$, $a\sqcup b=b \sqcup a$, $a\sqcap b = b \sqcap a$.

The triple $(Mj_1, Mj_2, Mn_3)$ is {\bf an MJN} on $M$ if $\forall x,y,z,\{Mj_1(x,y,z), Mj_2(x,y,z),$ $Mn_3(x,y,z)\} = \{x, y, z\}$ and for each triple
$(a, b, c) \in D^3$ with $\left\{a, b, c\right\}=\left\{x, y\right\}\in M$ operations $Mj_1(a, b, c)$, $Mj_2(a, b, c)$ return the unique majority element among $a, b, c$ (that occurs twice) and $Mn_3(a, b, c)$ returns the remaining minority element.
\end{definition}

The following theorem was established in \cite{kz13:jacm}.

\begin{theorem} \label{kolm} A conservative valued language $\bGamma$ is tractable if and only if there is a symmetric tournament polymorphism $(\sqcup,\sqcap)$ on $M$, an MJN $(Mj_1, Mj_2, Mn_3)$ on $\overline{M}=\left\{\left\{a,b\right\}|a,b\in D,a\ne b\right\}\setminus M$, such that $(\sqcup,\sqcap)$, $(Mj_1, Mj_2, Mn_3)\in \fPol\Gamma$.
\end{theorem}

Given $\bGamma=\left(D,f_1,...,f_s\right)$, let us construct a relational structure $\bGamma'_{c} = \left(D'_c,f'_{1},...,f'_{s}\right)$. Its domain, $D'_c$, is defined as a set of all triples 
$\big(M, (\sqcup,\sqcap), (Mj_1, Mj_2, Mn_3)\big)$ such that $(\sqcup,\sqcap)$ is a symmetric tournament polymorphism on $M$ and $(Mj_1, Mj_2, Mn_3)$ is an MJN on $\overline{M}$. All $f'_i$ will be relations, i.e. crisp cost functions.

A tuple 
$$
\big(\big(M^1, (\sqcup^1,\sqcap^1), (Mj^1_1, Mj^1_2, Mn^1_3)\big),\cdots,\big(M^p, (\sqcup^p,\sqcap^p), (Mj^p_1, Mj^p_2, Mn^p_3)\big)\big)$$ is in $f'_{i}$ if and only if $\big(\sqcup^1,\cdots, \sqcup^p\big)$, $\big(\sqcap^1,\cdots, \sqcap^p\big)$ and $\big(Mj^1_1,\cdots, Mj^p_1\big)$, $\big(Mj^1_2,\cdots, Mj^p_2\big)$, $\big(Mn^1_3,\cdots, Mn^p_3\big)$ are component-wise fractional polymorphisms of $f_i$, i.e. for any $\mathbf{x}=(x_{1},\cdots,x_{p})$,  $\mathbf{y}=(y_{1},\cdots,y_{p})$, $\mathbf{z}=(z_{1},\cdots,z_{p})$ the following inequalities are satisfied:
\begin{align*}
\vspace{-15pt}
f_i(\mathbf{x}\sqcup \mathbf{y}) + f_i(\mathbf{x}\sqcap \mathbf{y}) \le f_i(\mathbf{x})+f_i(\mathbf{y}) \\
f_i(Mj_1(\mathbf{x}, \mathbf{y}, \mathbf{z})) + f_i(Mj_2(\mathbf{x}, \mathbf{y}, \mathbf{z}))+f_i(Mn_3(\mathbf{x}, \mathbf{y}, \mathbf{z})) \le \\ f_i(\mathbf{x})+f_i(\mathbf{y})+f_i(\mathbf{z})
\vspace{-15pt}
\end{align*}
where $\mathbf{x}\sqcup \mathbf{y}=\big(x_1\sqcup^1 y_1, ..., x_{p}\sqcup^p y_{p}\big)$ and $\mathbf{x}\sqcap \mathbf{y}=\big(x_1\sqcap^1 y_1, ..., x_{p}\sqcap^p y_{p}\big)$. Analogously, $M(\mathbf{x}, \mathbf{y}, \mathbf{z}) = \big(M^1(x_1, y_1, z_1), ..., M^{p}(x_{p}, y_{p}, z_{p})\big)$, where instead of $M$ we can paste $Mj_1$, $Mj_2$, or $Mn_3$.

The structure $\bGamma'_{c}$ is an analog of $\bGamma'$. Its domain consists of fractional polymorphisms, that play the same role for valued CSPs as polymorphisms for the crisp case.

\begin{theorem} \label{gammashtrih} For conservative $\bGamma$, $\mathcal{H}^{\bGamma}_{c}=\up(\bGamma'_{c})$.
\end{theorem}
\ifTR
\begin{proof} [sketch] Using \eqref{max-cons} and theorem \ref{kolm} we conclude that
\begin{align*}
\mathcal{H}^{\bGamma}_{c}= \big\{\bR | \exists (\sqcup,\sqcap) (\textit{an STP on }M), (Mj_1, Mj_2, Mn_3) (\textit{MJN on }\overline{M}): \\
(\sqcup,\sqcap), (Mj_1, Mj_2, Mn_3)\in \ensuremath{fPol}(\bGamma_{\bR})\big\}
\end{align*}
Note that in the latter formula, $ (\sqcup,\sqcap)$ and $(Mj_1, Mj_2, Mn_3)$ are fractional polymorphisms defined on the domain $D_{\bR}$. 

Suppose that $\bR = (V, ...)\in \mathcal{H}^{\bGamma}_{c}$ and we are given $(\sqcup,\sqcap), (Mj_1, Mj_2, Mn_3)\in \ensuremath{fPol}(\bGamma_{\bR})$. Let us define a mapping $h: V\rightarrow D'_c$ by the following rule: each $v\in V$ is first mapped to a triple $\big(M|_{D_v}, (\sqcup|_{D_v},\sqcap|_{D_v}), (Mj_1|_{D_v}, Mj_2|_{D_v}, Mn_3|_{D_v})\big)$, where $M|_{D_v}=\left\{\left\{a,b\right\}\in M|a,b\in D_v\right\}$; at the second stage we identify elements $(v,a)\in D_v$ and $a\in D$ and obtain a resulting triple $h(v)$. It is easy to see that $h$ is a homomorphism from $\bR$ to $\bGamma'_c$. Thus, $\mathcal{H}^{\bGamma}_{c}\subseteq \up(\bGamma'_{c})$.

If, on the contrary, $o: \bR\rightarrow \bGamma'_c$ is a homomorphism, then we can construct $(\sqcup,\sqcap)$ that is an STP on some $M$, $(Mj_1, Mj_2, Mn_3)$ that is MJN on $\overline{M}$ such that
$(\sqcup,\sqcap), (Mj_1, Mj_2, Mn_3)\in \ensuremath{fPol}(\bGamma_{\bR})$. This can be done by the rule: for any $v\in V$, $\big(M|_{D_v}$, $(\sqcup|_{D_v},\sqcap|_{D_v})$, $(Mj_1|_{D_v}, Mj_2|_{D_v}$, $Mn_3|_{D_v})\big)$ (after identifying $(v,a)\in D_v$ and $a\in D$) should coinside with $o(v)$; on the cross domain arguments $(x,y,z)$ (i.e. $\forall v\in V, \left\{x,y,z\right\}\not\subseteq D_v$), operations $Mj_1, Mj_2, Mn_3$ are defined to be equal to $x$, $y$, $z$ respectively (in fact, in any way that does not violate conservativity);  on a set of cross-domain pairs $\{a,b\}, a\in D_u, b\in D_v, u\ne v$, $\sqcup,\sqcap$ are defined to be STP. It is easy to see that $(\sqcup,\sqcap)$, $(Mj_1, Mj_2, Mn_3)\in \ensuremath{fPol}(\bGamma_{\bR})$. Thus, $\up(\bGamma'_{c})\subseteq \mathcal{H}^{\bGamma}_{c}$.
\end{proof}
\else
\fi

\ifTR
Note that conservative VCSPs are defined over valued languages, whereas $\bGamma'_c$ is a crisp language.
Recall that in the previous case of CSPs we reduced $\textsc{CSP}(\bGamma)$ to $\textsc{CSP}(\bGamma')$. The key property that allowed us to do this was theorem \ref{below}. In the case of conservative VCSPs, analogous theorem does not hold, and $\textsc{VCSP}(\bGamma'_c)$ can be easier than $\textsc{VCSP}(\bGamma)$. 
The following example shows that.

\begin{example} Consider a valued template $\bGamma = \left(D, f, {\bf Un}_D\right)$ over $D=\{0,1\}$, where $f(x,y)=0$ if $(x,y)\in\{(0,0),(1,0),(0,1)\}$ and $f(1,1)=\infty$. In is easy to see that for any input 
$\bR=\left(V,E,U_1,...,U_k\right)$ an assignment $a: V\rightarrow D$ has a finite cost if and only if a set $\left\{v|a(v)=1\right\}$ is an independent set in a graph $(V,E)$. Therefore, conservative $\textsc{VCSP}(\bGamma)$ is equivalent to finding maximum weight independent set (with arbitrary weights of vertices). Let us prove that $\bGamma'_c\sim \left(D,\ne,D,D,...\right)$, i.e. $\up(\bGamma'_{c})$ is equal to a set of all inputs $\bR=\left(V,E,U_1,...,U_k\right)$ for which a graph $(V,E)$ is bipartite.

Indeed, if a graph $(V,E)$ is bipartite,  the problem can be solved by an algorithm for maximum weight independent set in bipartite graphs. Moreover, a set of inputs $\bR=\left(V,E,U_1,...,U_k\right)$ for which a graph $(V,E)$ is bipartite, is, obviously, up-closed and does not restrict unaries. Therefore such inputs are all in $\up(\bGamma'_{c})$, due to theorem \ref{gammashtrih}.

Now, suppose that $\bR=\left(V,E,...\right)$ is in $\up(\bGamma'_{c})$ and a graph $(V,E)$ is not bipartite, i.e. it contains an odd cycle $C_{2k+1}$. Since,  $\textsc{VCSP}_{\bGamma'_{c}}(\bGamma)$ is tractable, then $\textsc{VCSP}_{\bR}(\bGamma)$ is tractable, and therefore, maximum weight independent set in a graph that is homomorphic to an odd cycle $C_{2k+1}$ is a tractable problem. But the last problem is known to be NP-hard \cite{Takhanov10adichotomy}, therefore $\bR\notin\up(\bGamma'_{c})$

Thus, we proved that $\textsc{CSP}(\bGamma'_{c})$ is equivalent to checking whether $(V,E)$ is bipartite for an input $\bR=\left(V,E,U_1,...,U_k\right)$, and that is a tractable problem.
\end{example}

So, $\textsc{CSP}(\bGamma'_{c})$ can be easier than $\textsc{VCSP}(\bGamma)$. 
\else
\fi
\section{Some experiments and open problems}\label{exper}
We list here some experimental results and open problems
\begin{itemize}
\item In the case when $D=\{0,1\}$, it can be shown that in the definition of $\bGamma'$ Siggers pairs can be replaced with pairs $(g,w)$ where $g$ is unary and $w$ is a ternary weak near unanimity operation on $g(D)$ (the number of such pairs on $\{0,1\}$ is moderate). This allows a practical computation of $\bGamma'$s core. We experimented with random structures over the boolean domain ($\Gamma = \{\rho_1, \rho_2, \rho_3\}, \ar(\rho_i)\leq 3$) and found that the domain size of $\bGamma'$s core is never greater than 5. 
\item Since $\textsc{CSP}(\bGamma)$ is reducible to $\textsc{CSP}(\bGamma')$, an interesting problem is to find necessary and sufficient conditions for $\bGamma\sim \bGamma'$ (i.e. for the case when such reduction is trivial). Experiments showed that if $\Gamma=\{\rho\}, \rho\subseteq \{0,1\}^3$ is NP-hard, then $\bGamma\sim \bGamma'$. At the same time, if $\Gamma=\{\rho, \{0\}, \{1\}\}, \rho\subseteq \{0,1\}^3$ is NP-hard, then $\bGamma\not\sim \bGamma'$.
\item The number of Siggers pairs on $D$ grows as $O(|D|^{|D|^4})$ which does not allow the calculation of $\bGamma'$ even in the case when $|D|=3$. Upper bounds on the domain size of $\bGamma'$s core is an open problem. 
\item The problem of classifying all conservative $\bGamma$ for which $\textsc{CSP}(\bGamma'_{c})$ is tractable (modification: is solvable in Datalog \cite{Bodirsky:2006:DCS:2096664.2096717}) is also open. 
\item For a general valued template $\bGamma$ a construction of a structure analogous to $\bGamma'$ is complicated by the absence of a tractability characterization via fractional polymorphisms of fixed arity (characterization in \cite{Ochremiak14} deals with cyclic multimorphisms of any arity). The building of an analogous theory is a direction of future work. 
\item Are all crisp templates widely tractable, or is $\textsc{CSP}_{\bGamma'}(\bGamma)$ always tractable?
\end{itemize}







\bibliographystyle{plainurl}
\bibliography{lit}

\newcommand{\noopsort}[1]{}
\begin{thebibliography}{10}

\bibitem{BartoKozikLics10}
L.~Barto and M.~Kozik.
\newblock New conditions for {T}aylor varieties and {CSP}.
\newblock In {\em Proceedings of the 25th Annual {IEEE} Symposium on Logic in
  Computer Science, {LICS} 2010, 11-14 July 2010, Edinburgh, United Kingdom},
  pages 100--109, 2010.

\bibitem{Bodirsky:2006:DCS:2096664.2096717}
Manuel Bodirsky and V\'{\i}ctor Dalmau.
\newblock Datalog and constraint satisfaction with infinite templates.
\newblock In {\em Proceedings of the 23rd Annual Conference on Theoretical
  Aspects of Computer Science}, STACS'06, pages 646--659, Berlin, Heidelberg,
  2006. Springer-Verlag.

\bibitem{bulatov06:3-elementjacm}
A.~Bulatov.
\newblock A dichotomy theorem for constraint satisfaction problems on a
  3-element set.
\newblock {\em Journal of the ACM}, 53(1):66--120, 2006.

\bibitem{bulatov11:conservative}
A.~Bulatov.
\newblock Complexity of conservative constraint satisfaction problems.
\newblock {\em ACM Transactions on Computational Logic}, 12(4), 2011.
\newblock Article 24.

\bibitem{bulatov05:classifying}
A.~Bulatov, A.~Krokhin, and A.~Jeavons.
\newblock Classifying the {C}omplexity of {C}onstraints using {F}inite
  {A}lgebras.
\newblock {\em {SIAM} Journal on Computing}, 34(3):720--742, 2005.

\bibitem{Bulatov17a}
Andrei~A. Bulatov.
\newblock A dichotomy theorem for nonuniform csps.
\newblock {\em CoRR}, abs/1703.03021, 2017.

\bibitem{DBLP:conf/cp/BulinDJN13}
J.~Bul\'in, D.~Delic, M.~Jackson, and T.~Niven.
\newblock On the reduction of the {CSP} dichotomy conjecture to digraphs.
\newblock In Christian Schulte, editor, {\em CP}, volume 8124 of {\em Lecture
  Notes in Computer Science}, pages 184--199. Springer, 2013.

\bibitem{Cook:1971}
Stephen~A. Cook.
\newblock The complexity of theorem-proving procedures.
\newblock In {\em Proceedings of the Third Annual ACM Symposium on Theory of
  Computing}, STOC '71, pages 151--158, New York, NY, USA, 1971. ACM.

\bibitem{cz11:ai}
Martin~C. Cooper and Stanislav {\noopsort{ZZ}\v{Z}}ivn\'y.
\newblock Hybrid tractability of valued constraint problems.
\newblock {\em Artificial Intelligence}, 175(9-10):1555--1569, 2011.

\bibitem{ECCV_Best_paper}
Jia Deng, Nan Ding, Yangqing Jia, Andrea Frome, Kevin Murphy, Samy Bengio, Yuan
  Li, Hartmut Neven, and Hartwig Adam.
\newblock Large-scale object classification using label relation graphs.
\newblock In {\em Computer Vision - {ECCV} 2014 - 13th European Conference,
  Zurich, Switzerland, September 6-12, 2014, Proceedings, Part {I}}, pages
  48--64, 2014.

\bibitem{feder98:monotone}
Tom\'as Feder and Moshe~Y. Vardi.
\newblock The {C}omputational {S}tructure of {M}onotone {M}onadic {S{N}{P}} and
  {C}onstraint {S}atisfaction: {A} {S}tudy through {D}atalog and {G}roup
  {T}heory.
\newblock {\em {SIAM} Journal on Computing}, 28(1):57--104, 1998.

\bibitem{GREEN20081094}
Martin~J. Green and David~A. Cohen.
\newblock Domain permutation reduction for constraint satisfaction problems.
\newblock {\em Artificial Intelligence}, 172(8):1094 -- 1118, 2008.

\bibitem{LORA}
Gregory Gutin, Arash Rafiey, Anders Yeo, and Michael Tso.
\newblock Level of repair analysis and minimum cost homomorphisms of graphs.
\newblock {\em Discrete Applied Mathematics}, 154(6):881--889, 2006.

\bibitem{Hell}
Pavol Hell and Jaroslav Ne\v{s}et\v{r}il.
\newblock On the complexity of h-coloring.
\newblock {\em Journal of Combinatorial Theory, Series B}, 48(1):92 -- 110,
  1990.

\bibitem{jeavons14:beatcs}
P.~Jeavons, A.~Krokhin, and S.~\v{Z}ivn\'y.
\newblock The complexity of valued constraint satisfaction.
\newblock {\em Bulletin of the {EATCS}}, 113:21--55, 2014.

\bibitem{Jeavons:1998}
Peter Jeavons.
\newblock On the algebraic structure of combinatorial problems.
\newblock {\em Theor. Comput. Sci.}, 200(1-2):185--204, June 1998.

\bibitem{Jegou93a}
Philippe J\'egou.
\newblock Decomposition of domains based on the micro-structure of finite
  constraint-satisfaction problems.
\newblock In Richard Fikes and Wendy~G. Lehnert, editors, {\em AAAI}, pages
  731--736. AAAI Press / The MIT Press, 1993.

\bibitem{4ary}
Keith Kearnes, Petar Markovi\'c, and Ralph McKenzie.
\newblock {Optimal strong Mal'cev conditions for omitting type 1 in locally
  finite varieties.}
\newblock {\em Algebra Univers.}, 72(1):91--100, 2014.

\bibitem{sanjeev}
Sanjeev Khanna, Nathan Linial, and Shmuel Safra.
\newblock On the hardness of approximating the chromatic number.
\newblock {\em Combinatorica}, 20(3):393--415, 2000.

\bibitem{kz13:jacm}
V.~Kolmogorov and S.~{\noopsort{ZZ}\v{Z}}ivn\'y.
\newblock The complexity of conservative valued {C}{S}{P}s.
\newblock {\em Journal of the ACM}, 60(2), 2013.
\newblock Article 10.

\bibitem{kolmogorov15:mrt}
Vladimir Kolmogorov, Michal Rolinek, and Rustem Takhanov.
\newblock Effectiveness of structural restrictions for hybrid csps.
\newblock In {\em Proceedings of 26th International Symposium, (ISAAC 2015)},
  pages 566--577. Springer Berlin Heidelberg, 2015.

\bibitem{Kolmogorov:2002}
Vladimir Kolmogorov and Ramin Zabih.
\newblock What energy functions can be minimized via graph cuts?
\newblock In {\em Proceedings of the 7th European Conference on Computer
  Vision-Part III}, ECCV '02, pages 65--81, London, UK, UK, 2002.
  Springer-Verlag.

\bibitem{Ochremiak14}
Marcin Kozik and Joanna Ochremiak.
\newblock {\em Algebraic Properties of Valued Constraint Satisfaction Problem},
  pages 846--858.
\newblock Springer Berlin Heidelberg, Berlin, Heidelberg, 2015.

\bibitem{Maffray}
Frédéric Maffray and Myriam Preissmann.
\newblock On the np-completeness of the k-colorability problem for
  triangle-free graphs.
\newblock {\em Discrete Mathematics}, 162(1–3):313 -- 317, 1996.

\bibitem{MarotiMcKenzie}
M.~Mar\'oti and R.~McKenzie.
\newblock Existence theorems for weakly symmetric operations.
\newblock {\em Algebra universalis}, 59(3--4):463--489, October 2008.

\bibitem{DBLP:Rafiey}
Arash Rafiey, Jeff Kinne, and Tom{\'{a}}s Feder.
\newblock Dichotomy for digraph homomorphism problems.
\newblock {\em CoRR}, abs/1701.02409, 2017.

\bibitem{schaefer78:complexity}
T.~J. Schaefer.
\newblock The {C}omplexity of {S}atisfiability {P}roblems.
\newblock In {\em Proceedings of the 10th {A}nnual {A}{C}{M} {S}ymposium on
  {T}heory of {C}omputing ({S}{T}{O}{C}'78)}, pages 216--226. ACM, 1978.

\bibitem{Siggers}
M.~H. Siggers.
\newblock A strong {M}al'cev condition for locally finite varieties omitting
  the unary type.
\newblock {\em Algebra universalis}, 64(1--2):15--20, October 2010.

\bibitem{swarts}
Jacobus~Stephanus Swarts.
\newblock {\em The complexity of digraph homomorphisms: Local tournaments,
  injective homomorphisms and polymorphisms}.
\newblock PhD thesis, University of Victoria, Canada, 2008.

\bibitem{Takhanov10adichotomy}
Rustem~S. Takhanov.
\newblock A dichotomy theorem for the general minimum cost homomorphism
  problem.
\newblock In {\em In Proceedings of the 27th International Symposium on
  Theoretical Aspects of Computer Science (STACS)}, pages 657--668, 2010.

\bibitem{Thapper15:Sherali}
Johan Thapper and Stanislav Zivny.
\newblock Sherali-adams relaxations for valued csps.
\newblock In {\em Automata, Languages, and Programming - 42nd International
  Colloquium, {ICALP} 2015, Kyoto, Japan, July 6-10, 2015, Proceedings, Part
  {I}}, pages 1058--1069, 2015.

\bibitem{Zhuk17}
Dmitriy Zhuk.
\newblock The proof of {CSP} dichotomy conjecture.
\newblock {\em CoRR}, abs/1704.01914, 2017.

\end{thebibliography}

\end{document}